\documentclass[prl,twocolumn,floatfix,
amsmath,amssymb,aps,superscriptaddress,nofootinbib]{revtex4-1}
\usepackage{dcolumn,amsmath}
\usepackage[T2A]{fontenc}
\usepackage{graphicx}
\graphicspath{ {./Images/} }
\usepackage{bm}
\usepackage{xfrac}
\usepackage{indentfirst} 
\usepackage{physics}

\usepackage{xcolor}

\usepackage{hyperref}

\usepackage{amsmath}
\usepackage{graphicx}
\usepackage{bm}
\usepackage{xcolor}
\usepackage{multirow}
\usepackage{amssymb}
\usepackage{soul}

\newcommand{\eEDM}{{\em e}EDM}
\newcommand{\ecm}{\ensuremath{e {\cdotp} {\rm cm}}}

\newcommand{\de}{d_\mathrm{e}}

\newcommand{\B}{\mathcal{B}} 
\newcommand{\E}{\mathcal{E}} 

\begin{document}
\title{Accurate numerical evaluation of systematics in the experiment for electron electric dipole moment measurement in HfF$^+$}

\begin{abstract}
 Hyperfine structure of the ground rotational level of the metastable $^3\Delta_1$ electronic state of $^{180}$HfF$^+$ ion is calculated at presence of variable external electric and magnetic fields. Calculations are required for analysis of systematic effects in experiment for electron electric dipole moment ($e$EDM)  search. Different perturbations in molecular spectra important for $e$EDM spectroscopy are taken into account. 
\end{abstract}

\author{Alexander N. Petrov}
\email{petrov\_an@pnpi.nrcki.ru}
\affiliation{Petersburg Nuclear Physics Institute named by B.P.\ Konstantinov of National Research Center ``Kurchatov Institute'' (NRC ``Kurchatov Institute'' - PNPI), 1 Orlova roscha mcr., Gatchina, 188300 Leningrad region, Russia}
\affiliation{Saint Petersburg State University, 7/9 Universitetskaya nab., St. Petersburg, 199034 Russia}
\homepage{http://www.qchem.pnpi.spb.ru    }


\maketitle

\section{Introduction}
The measuring the electron electric dipole moment (\eEDM) serves as highly sensitive probe for testing the boundaries of the Standard Model of electroweak interactions and its extensions \cite{whitepaper, YamaguchiYamanaka2020,YamaguchiYamanaka2021}.
The current constrain for \eEDM\ $|\de|<4.1\times 10^{-30}$ \ecm\ (90\% confidence) was obtained using trapped $^{180}$Hf$^{19}$F$^+$ ions \cite{newlimit1} with spinless $^{180}$Hf isotope.
The measurements were performed on the ground rotational, $J{=}1$, level in the metastable electronic $^3\Delta_1$ state. As a matter of fact the \eEDM\ measurement is a highly accurate spectroscopy of $J{=}1$ level at the presence of rotating electric and magnetic fields.
It is clear that accurate evaluation of systematic
effects becomes very important with the increase in statistical sensitivity. The main part of a great success achieved in solving this problem in HfF$^+$ experiment is due to the existence of close levels, so-called $\Omega$-doublets, of the opposite parities.
In Ref. \cite{newlimit2} possible systematic shifts in the experiment were considered in details and corresponding analytical formulas were obtained. In turn in Refs. \cite{Petrov:17b, Petrov:18} the numerical method for theoretical calculation of $J{=}1$ hyperfine energy levels in rotating fields was developed. The method demonstrated a very high accuracy by comparison with the latest experimental data \cite{Petrov:23}. The goal of the present work is to study selected systematics numerically taking into account different perturbations in molecular spectra.

The \eEDM\ sensitive levels of $^{180}$Hf$^{19}$F$^+$ are described in details in Refs. \cite{Leanhardt:20111, Cornell:2017, newlimit2}. $^{180}$Hf isotope is spinless, $^{19}$F isotope has a non-zero nuclear spin $I{=}1/2$. Hyperfine energy splitting between levels with total momentum  $F=3/2$ and $F=1/2$, {\bf F}={\bf J}+{\bf I}, is several tens of megahertz. In the absence of external fields, each hyperfine level has two parity eigenstates known as the $\Omega$-doublets.
In the external {\it static} electric field the $F=3/2$ states form two (with absolute value of projection of the total momentum on the direction of the electric field, $m_F$, equal to one half and three half) Stark doublets levels. Below the levels in the doublets will be called upper and lower in accordance to their energies. Upper and lower levels in doublet are double degenerate. Namely, two Zeeman sublevels connected by time reversal $m_F \rightarrow -m_F$ have the same energy.
The levels $m_F=\pm3/2$ are of interest for the \eEDM\ search experiment. 
The corresponding energy scheme is depicted on Fig. 1 of Ref. \cite{Petrov:23}.

The picture above is for the static electric field. Now let us take into account the fact that the fields in the experiment are the rotating ones. 
The rotation of electric field causes the degenerate sublevels $m_F=+3/2$ and $m_F=-3/2$ to interact \cite{Leanhardt:20111}.
Therefore, in the case of {\it rotating} electric field eigenstates have slightly different energies and present equal-mixed combinations of $m_F=\pm3/2$ sublevels which are insensitive to \eEDM. Note, that in case of rotating electric field  $m_F$ is projection on the axis (coinciding with rotating electric field) rotating in the space.
In turn the rotating magnetic field which in the experiment is parallel or antiparallel to the rotating electric field gives opposite energy shift for $m_F=+3/2$ and $m_F=-3/2$ and for a sufficiently large magnetic field $m_F$ becomes a good quantum number (as in static fields) and corresponding eigenstates again become sensitive to \eEDM.
We see that magnetic field, in contrast to experiments in static fields, is not an (not only an) auxiliary tool, but should ensure a nonzero energy shift due to possible nonzero value of \eEDM\  \cite{Cornell:2017, Petrov:18}. To completely polarize the molecule and to access the maximum  \eEDM\ signal both rotating electric and magnetic fields should be large enough, see e.g. Fig. 2 in Ref. \cite{Petrov:18}. For these fields the energy splitting, $f$, between $m_F=\pm3/2$ sublevels is dominated by Zeeman interaction with smaller contribution coming from the fact that rotating fields are used. 

The measurement of $f$ is repeated under different conditions which depend on binary switch parameters such as $\tilde {\cal B}$, $\tilde {\cal D}$, $\tilde {\cal R}$ being switched from $+1$ to $-1$ (see Ref. \cite{Cornell:2017, newlimit1} for details).
$\tilde {\cal B} =+1(-1)$ means that the rotating magnetic field, ${\bf B}_{\rm rot}$, is parallel (antiparallel) to the rotating electric field ${\bf E}_{\rm rot}$; $\tilde {\cal D}=+1(-1)$ means that the measurement was performed for lower (upper)
Stark level; and $\tilde {\cal R}$ defines direction for the rotation of the fields around the laboratory $z$ axis:  ${\vec{ \omega}}_{\rm rot} = \tilde {\cal R}\omega_{\rm rot}\hat{z}$, where ${\vec{ \omega}}$ is the angular velocity.
The measured of $f$ can be expanded as 
\begin{eqnarray}
\nonumber
f({\cal \tilde{D}},{\cal \tilde{B}},{\cal \tilde{R}}) = f^{0} 
+{{\cal \tilde{D}}}f^{{\cal {D}}}
+{{\cal \tilde{B}}}f^{{\cal {B}}}
+{{\cal \tilde{R}}}f^{{\cal {R}}} \\
\label{fDBR}
+{{\cal \tilde{B}}{\cal \tilde{D}}}f^{{\cal {B}}{\cal {D}}} 
+{{\cal \tilde{D}}{\cal \tilde{R}}}f^{{\cal {D}}{\cal {R}}}
+{{\cal \tilde{B}}{\cal \tilde{R}}}f^{{\cal {B}}{\cal {R}}}
+{{\cal \tilde{D}}{\cal \tilde{B}}{\cal \tilde{R}}}f^{{\cal {D}}{\cal {B}}{\cal {R}}},
\end{eqnarray}
 where notation $f^{S_1,S_2...}$ denotes a component which is odd under the switches $S_1,S_2,...$ 
 and can be calculated by formula
\begin{equation}
f^{S_1,S_2...} = \frac{1}{8}\sum_{\tilde {\cal B}, \tilde {\cal D}, \tilde {\cal R}}{S_1S_2...}f({\cal \tilde{D}},{\cal \tilde{B}},{\cal \tilde{R}}).
\label{components}
\end{equation}

The \eEDM\ signal manifests as the contribution to $f^{ {\cal B} {\cal D}}$ channel according to 
\begin{equation}
  f^{ {\cal B} {\cal D}} = 2d_{e}E_{\rm eff},
\label{fbd}
\end{equation}
where $E_{\rm eff}$ is the effective electric field,
which can be obtained only in precise calculations of the electronic structure.
The values $E_{\rm eff}=$ 24~GV/cm \cite{Petrov:07a, Petrov:09b}, 22.5(0.9)~GV/cm \cite{Skripnikov:17c}, 22.7(1.4)~GV/cm \cite{Fleig:17} were obtained. 
According to eq. (\ref{components})
\begin{equation}
f^{{\cal {B}}{\cal {D}}} = \frac{1}{8}\sum_{\tilde {\cal B}, \tilde {\cal D}, \tilde {\cal R}}{\cal \tilde{B}}{\cal \tilde{D}}f({\cal \tilde{D}},{\cal \tilde{B}},{\cal \tilde{R}}).
\label{componentBD}
\end{equation}

Beyond \eEDM\ where are a lot of systematics which contributes to $f^{{\cal {B}}{\cal {D}}}$ and thus mimic \eEDM\ signal \cite{newlimit2}.
The point is that the measurement of other components $f^ {0}$ (even under all switches), $f^{ {\cal D}}$, $f^ {\cal B} $ and others together with their theoretical analysis can tell us about size of systematic effects and perhaps a way to take them into account \cite{newlimit2, Petrov:23}.

\section{Theoretical methods}
Following Refs. \cite{Petrov:11, Petrov:17b, Petrov:18}, the energy levels and wave functions of the  $^{180}$Hf$^{19}$F$^+$ ion are obtained by a numerical diagonalization of the molecular Hamiltonian (${\rm \bf \hat{H}}_{\rm mol}$) in the external variable electric ${\bf E}(\rm t)$ and magnetic ${\bf B}(\rm t)$ fields 
over the basis set of the electronic-rotational wavefunctions
\begin{equation}
 \Psi_{\Omega}\theta^{J}_{M,\Omega}(\alpha,\beta)U^{\rm F}_{M_I}.
\label{basis}
\end{equation}
Here $\Psi_{\Omega}$ is the electronic wavefunction, $\theta^{J}_{M,\Omega}(\alpha,\beta)=\sqrt{(2J+1)/{4\pi}}D^{J}_{M,\Omega}(\alpha,\beta,\gamma=0)$ is the rotational wavefunction, $\alpha,\beta,\gamma$ are Euler angles, $U^{F}_{M_I}$ is the F nuclear spin wavefunctions and $M$ $(\Omega)$ is the projection of the molecule angular momentum, {\bf J}, on the lab $\hat{z}$ (internuclear $\hat{n}$) axis, $M_I=\pm1/2$ is the projection of the nuclear angular 
momentum on the same axis. Note that $M_F=M_I+M$ is not equal to $m_F$. The latter, as stated above, is the projection of the total momentum on the rotating electric field.

The molecular Hamiltonian for $^{180}$Hf$^{19}$F$^+$ reads
\begin{equation}
{\rm \bf\hat{H}}_{\rm mol} = {\rm \bf \hat{H}}_{\rm el} + {\rm \bf \hat{H}}_{\rm rot} + {\rm \bf\hat{H}}_{\rm hfs} + {\rm \bf\hat{H}}_{\rm ext}.
\end{equation} 
Here ${\rm \bf \hat{H}}_{\rm el}$ is the electronic Hamiltonian, ${\rm \bf\hat{H}}_{\rm rot}$ is the Hamiltonian of the rotation of the molecule, ${\rm \bf\hat{H}}_{\rm hfs}$ is the hyperfine interaction between electrons and fluorine nuclei as they described in Ref. \cite{Petrov:17b}  and ${\rm \bf\hat{H}}_{\rm ext}$ describes the interaction of the molecule with variable magnetic and electric fields as it is described in Ref. \cite{Petrov:18}.

In this paper the time dependent electric and magnetic fields lie in the $xy$ plane.
Depending on the particular form of time dependence the interaction with the fields is taken into account within two approaches. In the first one the transition to the rotating frame is performed, whereas in the second approach the quantization of rotating electromagnetic field is performed.
Only the static fields parallel to ${\vec{ \omega}}_{\rm rot}$ ($\hat{z}$ axis)
are allowed in the first scheme, whereas the second approach is valid for arbitrary static, rotating and oscillating fields with arbitrary directions and frequencies \cite{Petrov:18}.

Following Ref. \cite{Petrov:17b} we considered $^3\Delta_1$,  $^3\Delta_2$,  $^3\Pi_{0^+}$ and $^3\Pi_{0^-}$ low-lying electronic basis states.
 ${\rm \bf \hat{H}}_{\rm el}$ is diagonal on the basis set (\ref{basis}). Its eigenvalues are  transition energies of these states. They were calculated and measured in Ref.~\cite{Cossel:12}:
\begin{align}
\label{Molbasis}
\nonumber
^3\Delta_1 & : T_e=976.930~{\rm cm}^{-1}\ ,\\
\nonumber
 ^3\Delta_2 & : T_e=2149.432~{\rm cm}^{-1}\ ,\\
 \nonumber
^3\Pi_{0^-} & : T_e=10212.623~{\rm cm}^{-1}\ ,\\
 ^3\Pi_{0^+} & : T_e=10401.723~{\rm cm}^{-1}\ .
\end{align}

 Electronic matrix elements for calculation of the molecular Hamiltonian were taken from Ref. \cite{Petrov:17b}, except for the hyperfine structure constant $A_{\parallel}= -62.0~{\rm MHz}$ measured in Ref. \cite{Cornell:2017}.

\section{Results}
\subsection{Non-reversing magnetic field}
 In the experiment the rotating magnetic field, ${\bf B}_{\rm rot}$, is parallel or antiparallel to the rotating electric field ${\bf E}_{\rm rot}$. In an ideal case after reversing the absolute value of magnetic field remains the same. At the presence of non-reversing component the absolute values for two directions are different. Non-reversing magnetic field makes additional contributions to $f^{ {\cal B} {\cal D}}$, which leads to systematic effect, as well as to  $f^{\cal B}$ components. Both shifts are proportional to non-reversing component of ${\bf B}_{\rm rot}$, and according to Ref. \cite{newlimit2} the ratio is
 \begin{equation}
  \frac{f^{\cal B}}{f^{ {\cal B} {\cal D}}}   = \frac{{\rm g}^u +{\rm g}^l}{{\rm g}^u -{\rm g}^l}.
  \label{ratio}
 \end{equation}
 Here ${\rm g}^u$ and ${\rm g}^l$ are the
 g-factors of the upper and lower Stark doublets in the external electric field. Thus, one can remove this systematic monitoring the relatively large $f^{\cal B}$ component and applying the correction to $f^{ {\cal B} {\cal D}}$ on the base of Eq. (\ref{ratio}).

For numerical calculation of this effect we, according to the first approach mentioned above, perform a transition to the rotating frame. In this case the rotating fields are replaced by the static ones in rotating frame:
\begin{eqnarray}
 {\bf E}(\rm t)_{\rm rot} = 
 \E_{\rm rot}(\hat{x} cos(\omega_{\rm rot}t) + \hat{y}sin(\omega_{\rm rot}t)) \rightarrow \E_{\rm rot}\hat{X},
\label{Erot0}
\end{eqnarray}
\begin{eqnarray}
 {\bf B}(\rm t)_{\rm rot} =  
 \B_{\rm rot}(\hat{x}cos(\omega_{\rm rot}t) + \hat{y}sin(\omega_{\rm rot}t)) \rightarrow  \B_{\rm rot}\hat{X}
\label{Brot0}
\end{eqnarray}
and the perturbation 
\begin{equation}
 \hat{V} = - \vec{ \omega}_{\rm rot}{\bf F} = -\omega_{\rm rot} \hat{F}_Z  
 \label{rotperp}
\end{equation}
due to the rotation is added to the Hamiltonian. Here $X$,$Y$,$Z$ are the axes of the rotating frame. 

\begin{figure}[h]
\centering
  \includegraphics[width=0.5\textwidth]{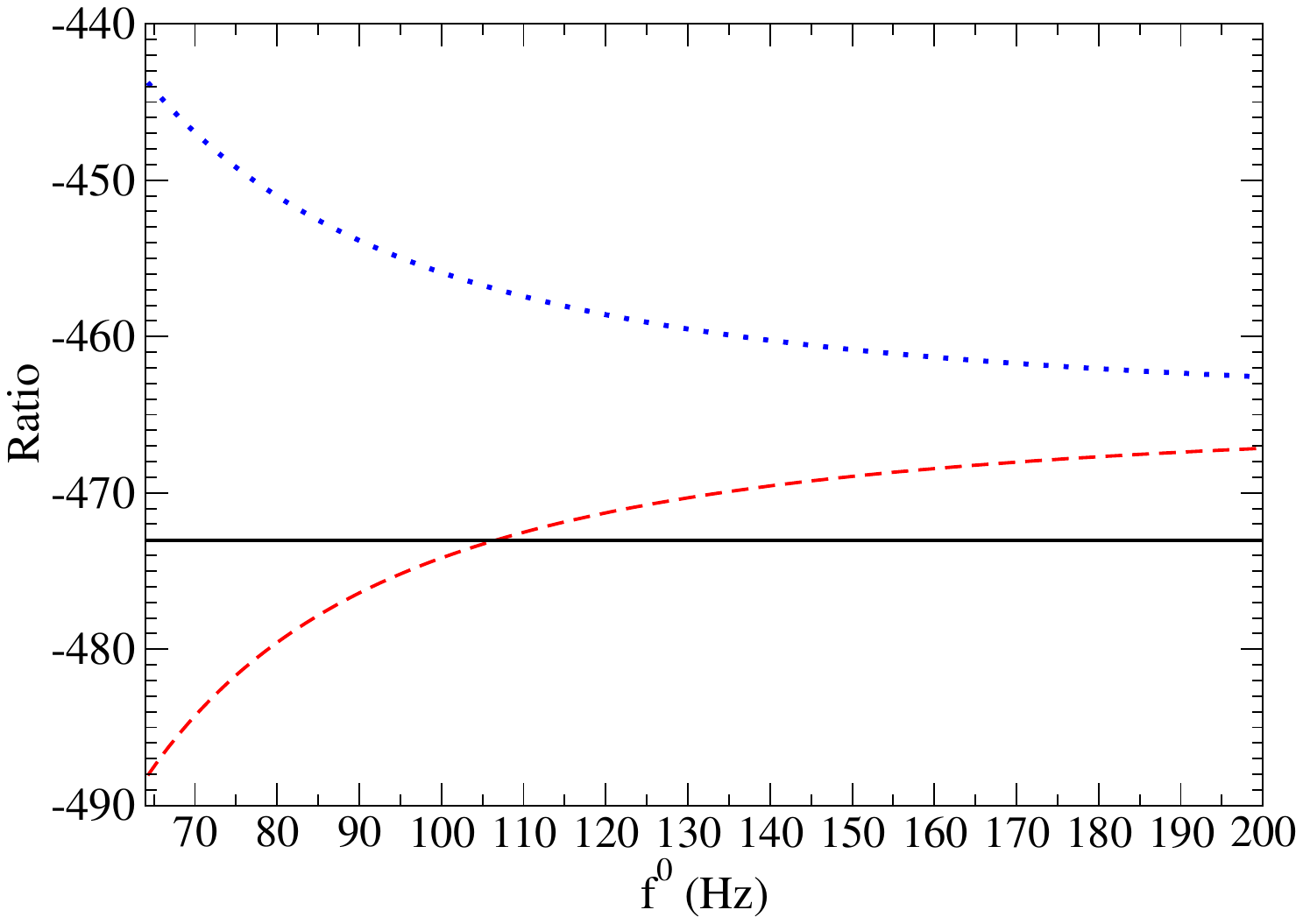}
  \caption{(Color online) Calculated ratios $f^{\cal B}/f^{ {\cal B} {\cal D}}$  (dashed(red) curve) and $f^{\cal D}/f^{ { 0}}$ (doted (blue) curve)
  as functions of the component$f^0$. Also the constant value $({\rm g}^u +{\rm g}^l)/({\rm g}^u -{\rm g}^l) = -473$ indicated as a black horizotal line.}
  \label{ratiofig}
\end{figure}

The calculated ratio  $f^{\cal B}/f^{ {\cal B} {\cal D}}$ 
as function of 
$f^0$ on Fig. \ref{ratiofig} is presented. In the calculation $\omega_{\rm rot}/2\pi = +375$ kHz and $\E_{\rm rot}=+58$ V/cm which correspond to the values used in the experiment. Also the calculated ratio $f^{\cal D}/f^{ {0}}$ and the calculated value $({\rm g}^u +{\rm g}^l)/({\rm g}^u -{\rm g}^l) = -473$ are given. For values $f^{ {0}} = $ 77 Hz, 105 Hz and 151 Hz, used in the experiment \cite{newlimit1}, we obtain $f^{\cal B}/f^{ {\cal B} {\cal D}}$ = $-481$, $-473$ and $-469$ respectively. The latter value corresponds to the solid (black) curve on Fig. 4 of Ref. \cite{Petrov:23}. The values are not identical to each other and to  
$({\rm g}^u +{\rm g}^l)/({\rm g}^u -{\rm g}^l)$ due to the rotation perturbation (\ref{rotperp}). As Zeeman splitting $f^{ {0}} $ increases the ratios  $f^{\cal B}/f^{ {\cal B} {\cal D}}$ and $f^{\cal D}/f^{ {0}}$ approach their saturated value $-465$ which is different from $({\rm g}^u +{\rm g}^l)/({\rm g}^u -{\rm g}^l) = -473$ on 8.

 \subsection{The second and higher harmonics of $\E_{\rm rot}$}
According to the theory of Ref. \cite{newlimit2} addition electric field oscillating in $xy$ plane at double frequency $2\omega_{\rm rot}$ together with static magnetic field in the same plane makes additional contributions to $f^ {\cal B} $ but no contribution to $f^{ {\cal B} {\cal D}}$ which formally does not lead to a systematic effect.
However, applying the correction (\ref{ratio}) on the base of the observed $f^ {\cal B} $ does affect the measurement of  $f^{ {\cal B} {\cal D}}$ component \cite{newlimit2}.
 
To calculate this effect we use variable fields which in addition to the components that rotates in the $xy$-plane with frequency $\omega_{\rm rot}$ (see Eqs. (\ref{Erot0},\ref{Brot0})) consists of static component of magnetic field along the laboratory $x$ axis and electric field with components along $x$ and $y$ axes which oscillate with frequency $2\omega_{\rm rot}$ and have addition to rotating component phase $\varphi$:

\begin{eqnarray}
\nonumber
 {\bf E}(\rm t) =  \E_{\rm rot}(\hat{x} cos(\omega_{\rm rot}t) + \hat{y}sin(\omega_{\rm rot}t)) + \\
 \E_{\rm x}\hat{x} cos(2\omega_{\rm rot}t+\varphi) + \E_{\rm y}\hat{y} cos(2\omega_{\rm rot}t+\varphi),
\label{Erot}
\end{eqnarray}
\begin{eqnarray}
 {\bf B}(\rm t) =  B_x\hat{x} +  \\
 \B_{\rm rot}(\hat{x}cos(\omega_{\rm rot}t) + \hat{y}sin(\omega_{\rm rot}t)).
\label{Brot}
\end{eqnarray}
Below we put $\omega_{\rm rot}/2\pi = +375$ kHz, $\E_{\rm rot}=+58$ V/cm, $\B_{\rm rot}=\pm 6$ mG (corresponds to $f^0=77$ MHz) which are the values used in the experiment \cite{newlimit1} and $B_x= 14$ mG and 
$\E_x = \E_y $ , $\E_x / \E_{\rm rot} $ = 10$^{-2}$.
Note, that $\omega_{\rm rot}$ and $\E_{\rm rot}$ are always positive.
In this and following subsections the time-dependence of external fields
is accounted for by the interaction with the corresponding quantized electromagnetic fields that corresponds to the second approach described in Ref. \cite{Petrov:18}.

\begin{figure}[h]
\centering
  \includegraphics[width=0.5\textwidth]{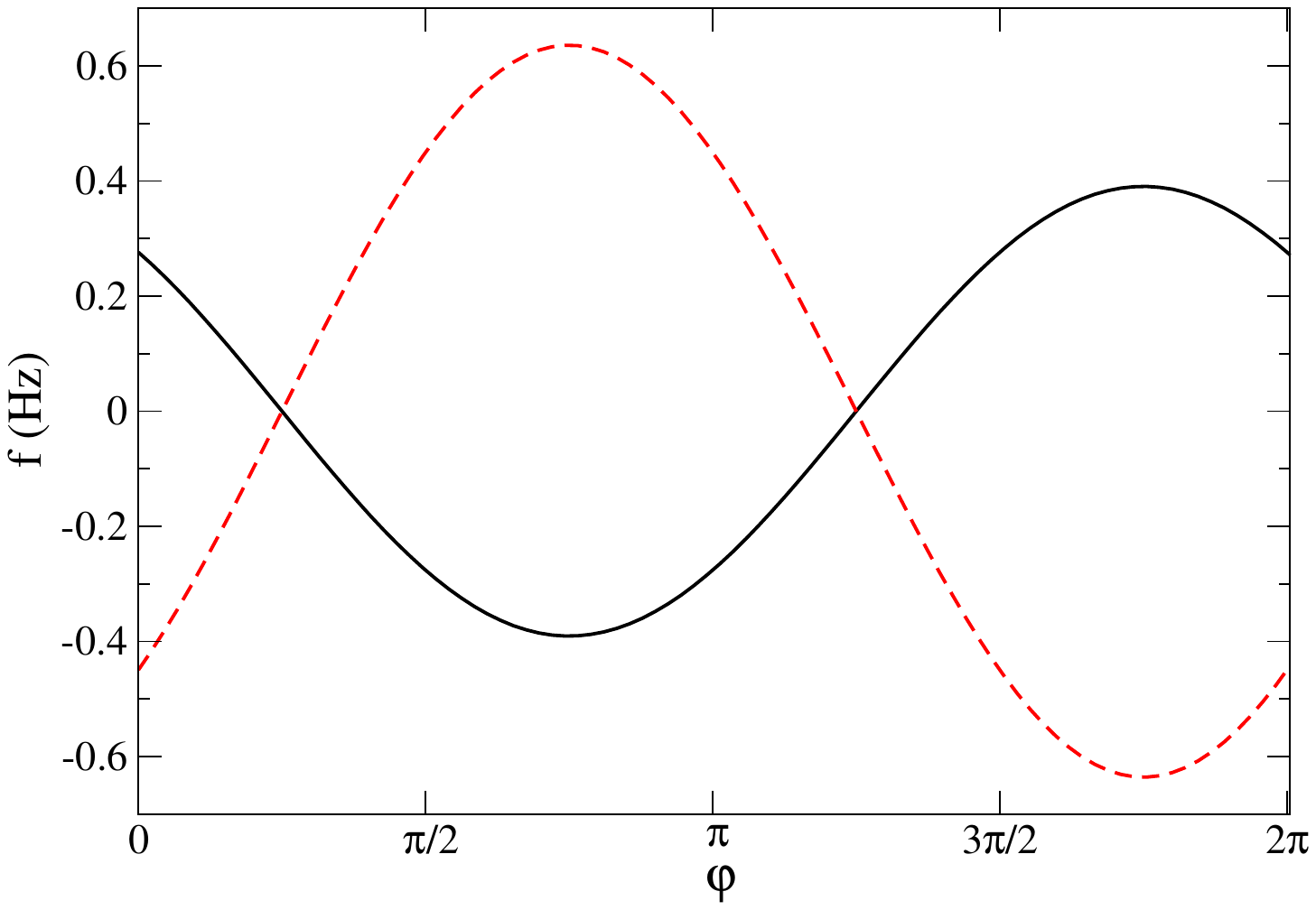}
  \caption{(Color online) Calculated $10^4f^{ {\cal B} {\cal D}}$ (solid (black) curve) and $f^{ {\cal B} }$ (dashed (red) curve)
  as functions of the phase $\varphi$.}
  \label{fBNfB}
\end{figure}

In Fig. \ref{fBNfB} the calculated values of $f^{ {\cal B} {\cal D}}$ and $f^{ {\cal B} }$ as functions of the phase $\varphi$ are given. The calculated $f^{ {\cal B} }$ is in agreement with Fig 3 Panel B of Ref. \cite{newlimit1}. The general behavior with presence  of static magnetic field  along the $y$ axis is given by eq. (37) of Ref. \cite{newlimit2}. Our calculation also indicates nonzero value 
 $f^{ {\cal B} {\cal D}}$ with the ratio  $f^{\cal B}/f^{ {\cal B} {\cal D}}=-16000$.
 
\begin{figure}[h]
\centering
  \includegraphics[width=0.5\textwidth]{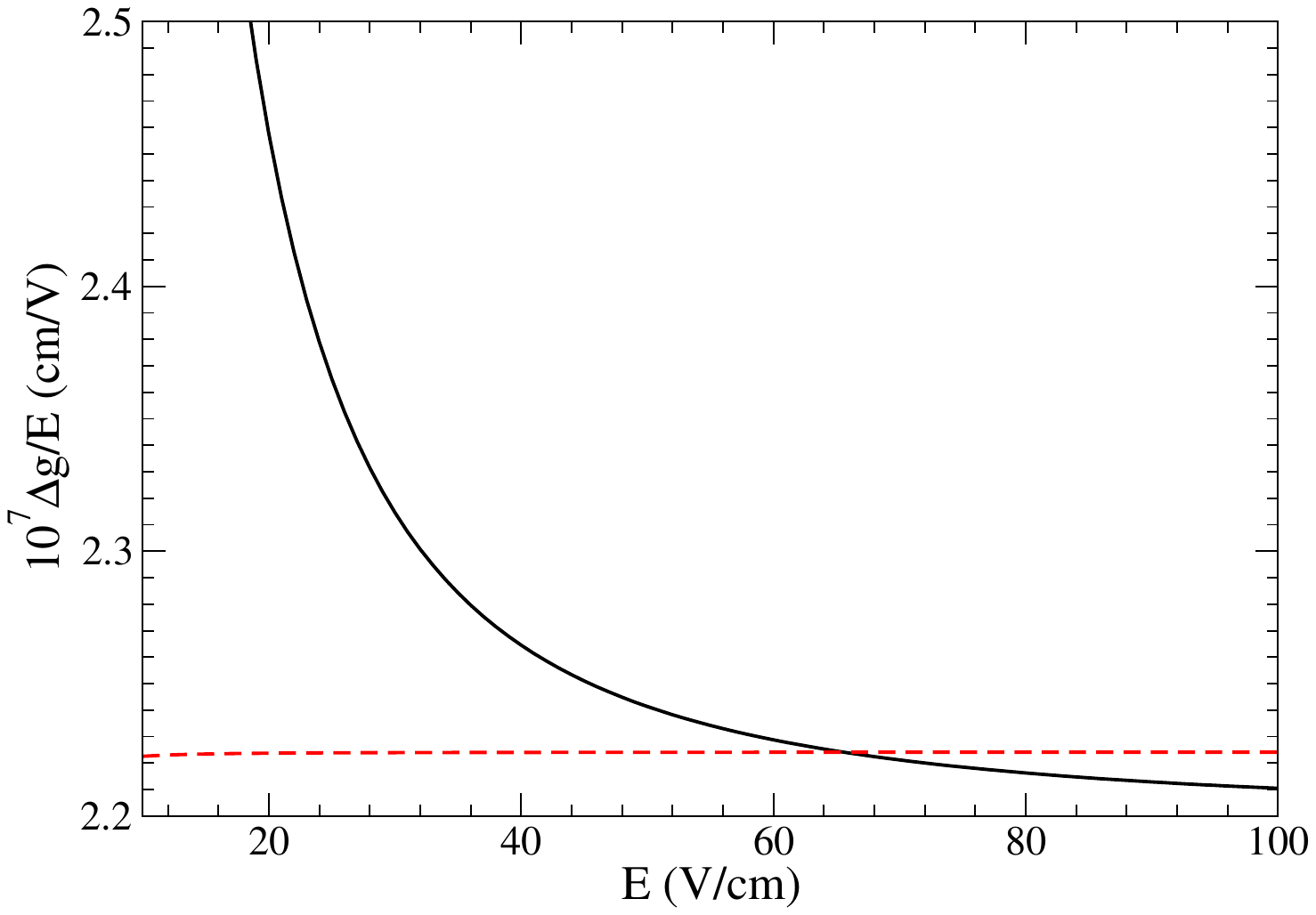}
  \caption{(Color online) Calculated $\Delta {\rm g}/E$ as function of the external {\it static} electric field $E$. Solid (black) curve --  magnetic interaction with both $^3\Pi_{0^\pm}$ and $^3\Delta_2$ is taken into account. Dashed (red) curve -- the interaction with $^3\Pi_{0^\pm}$ is omitted.}
  \label{deltag}
\end{figure}

\begin{table}
\caption{ The calculated $\Delta{\rm g}_0$ and  $\Delta {\rm g}_1$ (cm/V)}
\label{deltag01}
\begin{tabular}{ccc}
E(V/cm)                     &  $10^7 \Delta{\rm g}_0$ & $10^7\Delta {\rm g}_1$ \\
\hline
   10 & 19.6 & 1.27  \\
   20 &10.1 & 1.95  \\
   30 & 6.8 & 2.09  \\
   40 & 5.1 & 2.14  \\
   50 & 4.1 & 2.16  \\
   60 & 3.4 & 2.17  \\
   70 & 2.9 & 2.18  \\
   80 & 2.6 & 2.18  \\
   90 & 2.3 & 2.19  \\
  100 & 2.0 & 2.19  
\end{tabular}
\end{table}

    According to the theory of Ref. \cite{newlimit2} the nonzero value of $f^{ {\cal B} {\cal D}}$ can appear if $\Delta {\rm g}/E$ depends on the external {\it static} electric field $E$. Here $\Delta {\rm g} = {\rm g}^u - {\rm g}^l$.
    Fig.~\ref{deltag} presents the calculated values of  $\Delta {\rm g}/E$. We present results for the cases when magnetic interaction with both $^3\Pi_{0^\pm}$ and $^3\Delta_2$ is taken into account and for the case when the interaction with $^3\Pi_{0^\pm}$ is omitted. One can see that if interaction with $^3\Pi_{0^\pm}$ states is taken into account the value $\Delta {\rm g}/E$ depends on the external electric fields. Within small area of an value of static electric field the g-factor difference can be presented as 
    \begin{equation}
    \Delta {\rm g} = \Delta {\rm g}_0 +  \Delta {\rm g}_1 E. 
    \end{equation}
    If interaction with $^3\Pi_{0^\pm}$ is omitted $\Delta {\rm g}_0 = 0$ for a very high accuracy.
    In table \ref{deltag01} the calculated $\Delta{\rm g}_0$ and  $\Delta {\rm g}_1$ for the case when interaction with $^3\Pi_{0^\pm}$ is taken into account are given. Interaction with $^3\Pi_{0^\pm}$ states ensure nonzero $\Delta {\rm g}$ value for $\Omega-$doublets levels already at zero external electric field \cite{Petrov:17b}. Note, that one of the $\Omega-$doublet states has admixture only $^3\Pi_{0^+}$ state, whereas another one has admixture only $^3\Pi_{0^-}$. As electric field increases $\Omega-$doublets levels become the Stark-doublets ones with good $\Omega$ quantum number and with equal admixture of $^3\Pi_{0^+}$ as well as $^3\Pi_{0^-}$. Therefore as electric field increases the  $\Delta {\rm g}_0$ decreases, but is nonzero for any finite electric field. As is stated above, the nonzero  $\Delta {\rm g}_0$ leads to nonzero  $f^{ {\cal B} {\cal D}}$. As the effect is proportional to $\Delta {\rm g}_0$, according to the Table \ref{deltag01}, for $\E_{\rm rot} $ = 20 V/cm used in the first stage of the experiment \cite{Cornell:2017} we have $f^{\cal B}/f^{ {\cal B} {\cal D}}=-5500$. 

Similarly to the second harmonic the electric field oscillating in $xy$ plane at frequency $3\omega_{\rm rot}$ together with the gradient of magnetic field in the same plane makes additional contributions to both $f^{\cal B}$ and $f^{ {\cal B} {\cal D}}$ with the same (as for the second harmonic) ratio $f^{\cal B} / f^{ {\cal B} {\cal D}} = \Delta g_0/(g^u + g^l)$. The absolute value for $f^{\cal B}$ is given by Eq. (39) of Ref. \cite{newlimit2}.

\subsection{Ellipticity of $\E_{\rm rot}$}
According to the theory of Ref. \cite{newlimit2} ellipticity of $\E_{\rm rot}$ together with first-order magnetic field gradient
 makes additional contributions to $f^{ {\cal B} {\cal D}}$ and $f^{\cal B}$ components with the ratio
 \begin{equation}
  \frac{f^{\cal B}}{f^{ {\cal B} {\cal D}}}   = \frac{3}{4}\frac{{\rm g}^u +{\rm g}^l}{{\rm g}^u -{\rm g}^l}.
  \label{ratioe}
 \end{equation}
To calculate this effect we use variable fields
\begin{eqnarray}
\nonumber
 {\bf E}(\rm t) =  
 (\E_{\rm rot}+\E_\epsilon)\hat{x} cos(\omega_{\rm rot}t) + \\(\E_{\rm rot}-\E_\epsilon)\hat{y}sin(\omega_{\rm rot}t),
\label{Erote}
\end{eqnarray}
\begin{eqnarray}
\nonumber
 {\bf B}(\rm t) =
 \B_{\rm rot}\frac{\E_{\rm rot}+\E_\epsilon}{\E_{\rm rot}}\hat{x}cos(\omega_{\rm rot}t) + \\
 \nonumber
 \B_{\rm rot}\frac{\E_{\rm rot}-\E_\epsilon}{\E_{\rm rot}}\hat{y}sin(\omega_{\rm rot}t)) + \\
 \nonumber
 \B_{\epsilon}\frac{\E_{\rm rot}+\E_\epsilon}{\E_{\rm rot}}\hat{x}cos(\omega_{\rm rot}t) - \\
 \B_{\epsilon}\frac{\E_{\rm rot}-\E_\epsilon}{\E_{\rm rot}}\hat{y}sin(\omega_{\rm rot}t)). 
\label{Brote}
\end{eqnarray}
 In the calculation we put $\E_{\epsilon}=+1$ V/cm, $\B_{\epsilon}=+ 0.1$ mG.
 Equation (\ref{Erote}) is rotating electric field having an ellipticity with major axis along the $x$ axis. First two lines of Eq. (\ref{Brote}) is the modification of the rotating magnetic field from Eq. (\ref{Brot0}) caused by perturbation of the ion micromotion due to the acquired ellipticity of $\E_{\rm rot}$. This modification, actually, does not affect the result.  The last two lines of Eq. (\ref{Brote}) is the additional magnetic field feeling by the ion in the  first-order magnetic field gradient \cite{newlimit2}. 
 
 The calculation gives $f^{ {\cal B} {\cal D}} = -0.913\cdot 10^{-4}$ Hz,  $f^ {\cal B} = 0.332\cdot 10^{-1}$ Hz. The ratio is
 \begin{equation}
  \frac{f^{\cal B}}{f^{ {\cal B} {\cal D}}}   = -364 = 0.757 \cdot (-481),
  \label{ratioecalc}
 \end{equation}
where $-481$  is the ratio $f^{\cal B}/f^{ {\cal B} {\cal D}}$ for systematic related to the non-reversing magnetic field for $\B_{\rm rot}=\pm 6$ mG ($f^{0}$ = 77 Hz). Note the difference of the coefficient 0.757 in Eq. (\ref{ratioecalc}) from the coefficient 3/4 = 0.750 in Eq. (\ref{ratioe}). This difference can be explained as following. Looking on the derivation of Eq. (\ref{ratioe}) (see Eqs. (43,44) in Ref. \cite{newlimit2}) one notes that the coefficient 3/4 originate from the fact that ${\rm g}^u +{\rm g}^l$ is assumed to be independent of electric field, whereas
$\Delta{\rm g}={{\rm g}^u -{\rm g}^l}$ linearly depends on electric field. If $\Delta {\rm g}={{\rm g}^u -{\rm g}^l}$ were independent of electric field the coefficient in Eq. (\ref{ratioe}) would be equal to one. We know, however, from the calculation above that $\Delta {\rm g}$ has a small fraction (2.7\% for $\E_{\rm rot}$ =58 V/cm as it follows from Table \ref{deltag01}) which is independent of electric field. Then one can calculate that $1\cdot0.027 + 0.750(1-0.027)=0.757$ in accordance to the coefficient in Eq. (\ref{ratioecalc}).

\section{Conclusion}
The accurate numerical calculation of some systematic effects in the experiment for \eEDM\ search on $^{180}$HfF$^+$ cation is performed. A small deviation from analytical formulas derived in Ref. \cite{newlimit2} is discussed. The results can be used for testing experimental methods and in the next generation of experiments on the HfF$^+$ cation and on similar systems like ThF$^+$.

\bibliographystyle{apsrev}


\begin{thebibliography}{16}
\expandafter\ifx\csname natexlab\endcsname\relax\def\natexlab#1{#1}\fi
\expandafter\ifx\csname bibnamefont\endcsname\relax
  \def\bibnamefont#1{#1}\fi
\expandafter\ifx\csname bibfnamefont\endcsname\relax
  \def\bibfnamefont#1{#1}\fi
\expandafter\ifx\csname citenamefont\endcsname\relax
  \def\citenamefont#1{#1}\fi
\expandafter\ifx\csname url\endcsname\relax
  \def\url#1{\texttt{#1}}\fi
\expandafter\ifx\csname urlprefix\endcsname\relax\def\urlprefix{URL }\fi
\providecommand{\bibinfo}[2]{#2}
\providecommand{\eprint}[2][]{\url{#2}}

\bibitem[{\citenamefont{Alarcon et~al.}(2022)\citenamefont{Alarcon, Alexander,
  Anastassopoulos, Aoki, Baartman, BaeГџler, Bartoszek, Beck, Bedeschi, Berger
  et~al.}}]{whitepaper}
\bibinfo{author}{\bibfnamefont{R.}~\bibnamefont{Alarcon}},
  \bibinfo{author}{\bibfnamefont{J.}~\bibnamefont{Alexander}},
  \bibinfo{author}{\bibfnamefont{V.}~\bibnamefont{Anastassopoulos}},
  \bibinfo{author}{\bibfnamefont{T.}~\bibnamefont{Aoki}},
  \bibinfo{author}{\bibfnamefont{R.}~\bibnamefont{Baartman}},
  \bibinfo{author}{\bibfnamefont{S.}~\bibnamefont{BaeГџler}},
  \bibinfo{author}{\bibfnamefont{L.}~\bibnamefont{Bartoszek}},
  \bibinfo{author}{\bibfnamefont{D.~H.} \bibnamefont{Beck}},
  \bibinfo{author}{\bibfnamefont{F.}~\bibnamefont{Bedeschi}},
  \bibinfo{author}{\bibfnamefont{R.}~\bibnamefont{Berger}},
  \bibnamefont{et~al.}, \emph{\bibinfo{title}{Electric dipole moments and the
  search for new physics}} (\bibinfo{year}{2022}),
  \urlprefix\url{https://arxiv.org/abs/2203.08103}.

\bibitem[{\citenamefont{Yamaguchi and Yamanaka}(2020)}]{YamaguchiYamanaka2020}
\bibinfo{author}{\bibfnamefont{Y.}~\bibnamefont{Yamaguchi}} \bibnamefont{and}
  \bibinfo{author}{\bibfnamefont{N.}~\bibnamefont{Yamanaka}},
  \bibinfo{journal}{Phys. Rev. Lett.} \textbf{\bibinfo{volume}{125}},
  \bibinfo{pages}{241802} (\bibinfo{year}{2020}), \eprint{2003.08195}.

\bibitem[{\citenamefont{Yamaguchi and Yamanaka}(2021)}]{YamaguchiYamanaka2021}
\bibinfo{author}{\bibfnamefont{Y.}~\bibnamefont{Yamaguchi}} \bibnamefont{and}
  \bibinfo{author}{\bibfnamefont{N.}~\bibnamefont{Yamanaka}},
  \bibinfo{journal}{Phys. Rev. D} \textbf{\bibinfo{volume}{103}},
  \bibinfo{pages}{013001} (\bibinfo{year}{2021}), \eprint{2006.00281}.

\bibitem[{\citenamefont{Roussy et~al.}(2023)\citenamefont{Roussy, Caldwell,
  Wright, Cairncross, Shagam, Ng, Schlossberger, Park, Wang, Ye
  et~al.}}]{newlimit1}
\bibinfo{author}{\bibfnamefont{T.~S.} \bibnamefont{Roussy}},
  \bibinfo{author}{\bibfnamefont{L.}~\bibnamefont{Caldwell}},
  \bibinfo{author}{\bibfnamefont{T.}~\bibnamefont{Wright}},
  \bibinfo{author}{\bibfnamefont{W.~B.} \bibnamefont{Cairncross}},
  \bibinfo{author}{\bibfnamefont{Y.}~\bibnamefont{Shagam}},
  \bibinfo{author}{\bibfnamefont{K.~B.} \bibnamefont{Ng}},
  \bibinfo{author}{\bibfnamefont{N.}~\bibnamefont{Schlossberger}},
  \bibinfo{author}{\bibfnamefont{S.~Y.} \bibnamefont{Park}},
  \bibinfo{author}{\bibfnamefont{A.}~\bibnamefont{Wang}},
  \bibinfo{author}{\bibfnamefont{J.}~\bibnamefont{Ye}}, \bibnamefont{et~al.},
  \bibinfo{journal}{Science} \textbf{\bibinfo{volume}{381}},
  \bibinfo{pages}{46} (\bibinfo{year}{2023}),
  \eprint{https://www.science.org/doi/pdf/10.1126/science.adg4084},
  \urlprefix\url{https://www.science.org/doi/abs/10.1126/science.adg4084}.

\bibitem[{\citenamefont{Caldwell et~al.}(2023)\citenamefont{Caldwell, Roussy,
  Wright, Cairncross, Shagam, Ng, Schlossberger, Park, Wang, Ye
  et~al.}}]{newlimit2}
\bibinfo{author}{\bibfnamefont{L.}~\bibnamefont{Caldwell}},
  \bibinfo{author}{\bibfnamefont{T.~S.} \bibnamefont{Roussy}},
  \bibinfo{author}{\bibfnamefont{T.}~\bibnamefont{Wright}},
  \bibinfo{author}{\bibfnamefont{W.~B.} \bibnamefont{Cairncross}},
  \bibinfo{author}{\bibfnamefont{Y.}~\bibnamefont{Shagam}},
  \bibinfo{author}{\bibfnamefont{K.~B.} \bibnamefont{Ng}},
  \bibinfo{author}{\bibfnamefont{N.}~\bibnamefont{Schlossberger}},
  \bibinfo{author}{\bibfnamefont{S.~Y.} \bibnamefont{Park}},
  \bibinfo{author}{\bibfnamefont{A.}~\bibnamefont{Wang}},
  \bibinfo{author}{\bibfnamefont{J.}~\bibnamefont{Ye}}, \bibnamefont{et~al.},
  \bibinfo{journal}{Phys. Rev. A} \textbf{\bibinfo{volume}{108}},
  \bibinfo{pages}{012804} (\bibinfo{year}{2023}),
  \urlprefix\url{https://link.aps.org/doi/10.1103/PhysRevA.108.012804}.

\bibitem[{\citenamefont{Petrov et~al.}(2017)\citenamefont{Petrov, Skripnikov,
  and Titov}}]{Petrov:17b}
\bibinfo{author}{\bibfnamefont{A.~N.} \bibnamefont{Petrov}},
  \bibinfo{author}{\bibfnamefont{L.~V.} \bibnamefont{Skripnikov}},
  \bibnamefont{and} \bibinfo{author}{\bibfnamefont{A.~V.} \bibnamefont{Titov}},
  \bibinfo{journal}{Phys. Rev. A} \textbf{\bibinfo{volume}{96}},
  \bibinfo{pages}{022508} (\bibinfo{year}{2017}).

\bibitem[{\citenamefont{Petrov}(2018)}]{Petrov:18}
\bibinfo{author}{\bibfnamefont{A.~N.} \bibnamefont{Petrov}},
  \bibinfo{journal}{Phys. Rev. A} \textbf{\bibinfo{volume}{97}},
  \bibinfo{pages}{052504} (\bibinfo{year}{2018}).

\bibitem[{\citenamefont{Petrov et~al.}(2023)\citenamefont{Petrov, Skripnikov,
  and Titov}}]{Petrov:23}
\bibinfo{author}{\bibfnamefont{A.~N.} \bibnamefont{Petrov}},
  \bibinfo{author}{\bibfnamefont{L.~V.} \bibnamefont{Skripnikov}},
  \bibnamefont{and} \bibinfo{author}{\bibfnamefont{A.~V.} \bibnamefont{Titov}},
  \bibinfo{journal}{Phys. Rev. A} \textbf{\bibinfo{volume}{107}},
  \bibinfo{pages}{062814} (\bibinfo{year}{2023}),
  \urlprefix\url{https://link.aps.org/doi/10.1103/PhysRevA.107.062814}.

\bibitem[{\citenamefont{Leanhardt et~al.}(2011)\citenamefont{Leanhardt, Bohn,
  Loh, Maletinsky, Meyer, Sinclair, Stutz, and Cornell}}]{Leanhardt:20111}
\bibinfo{author}{\bibfnamefont{A.}~\bibnamefont{Leanhardt}},
  \bibinfo{author}{\bibfnamefont{J.}~\bibnamefont{Bohn}},
  \bibinfo{author}{\bibfnamefont{H.}~\bibnamefont{Loh}},
  \bibinfo{author}{\bibfnamefont{P.}~\bibnamefont{Maletinsky}},
  \bibinfo{author}{\bibfnamefont{E.}~\bibnamefont{Meyer}},
  \bibinfo{author}{\bibfnamefont{L.}~\bibnamefont{Sinclair}},
  \bibinfo{author}{\bibfnamefont{R.}~\bibnamefont{Stutz}}, \bibnamefont{and}
  \bibinfo{author}{\bibfnamefont{E.}~\bibnamefont{Cornell}},
  \bibinfo{journal}{Journal of Molecular Spectroscopy}
  \textbf{\bibinfo{volume}{270}}, \bibinfo{pages}{1 } (\bibinfo{year}{2011}),
  ISSN \bibinfo{issn}{0022-2852},
  \urlprefix\url{http://www.sciencedirect.com/science/article/pii/S0022285211001718}.

\bibitem[{\citenamefont{Cairncross et~al.}(2017)\citenamefont{Cairncross,
  Gresh, Grau, Cossel, Roussy, Ni, Zhou, Ye, and Cornell}}]{Cornell:2017}
\bibinfo{author}{\bibfnamefont{W.~B.} \bibnamefont{Cairncross}},
  \bibinfo{author}{\bibfnamefont{D.~N.} \bibnamefont{Gresh}},
  \bibinfo{author}{\bibfnamefont{M.}~\bibnamefont{Grau}},
  \bibinfo{author}{\bibfnamefont{K.~C.} \bibnamefont{Cossel}},
  \bibinfo{author}{\bibfnamefont{T.~S.} \bibnamefont{Roussy}},
  \bibinfo{author}{\bibfnamefont{Y.}~\bibnamefont{Ni}},
  \bibinfo{author}{\bibfnamefont{Y.}~\bibnamefont{Zhou}},
  \bibinfo{author}{\bibfnamefont{J.}~\bibnamefont{Ye}}, \bibnamefont{and}
  \bibinfo{author}{\bibfnamefont{E.~A.} \bibnamefont{Cornell}},
  \bibinfo{journal}{Phys.\ Rev.\ Lett.} \textbf{\bibinfo{volume}{119}},
  \bibinfo{pages}{153001} (\bibinfo{year}{2017}).

\bibitem[{\citenamefont{Petrov et~al.}(2007)\citenamefont{Petrov, Mosyagin,
  Isaev, and Titov}}]{Petrov:07a}
\bibinfo{author}{\bibfnamefont{A.~N.} \bibnamefont{Petrov}},
  \bibinfo{author}{\bibfnamefont{N.~S.} \bibnamefont{Mosyagin}},
  \bibinfo{author}{\bibfnamefont{T.~A.} \bibnamefont{Isaev}}, \bibnamefont{and}
  \bibinfo{author}{\bibfnamefont{A.~V.} \bibnamefont{Titov}},
  \bibinfo{journal}{Phys.\ Rev.\ A} \textbf{\bibinfo{volume}{76}},
  \bibinfo{pages}{030501(R)} (\bibinfo{year}{2007}).

\bibitem[{\citenamefont{Petrov et~al.}(2009)\citenamefont{Petrov, Mosyagin, and
  Titov}}]{Petrov:09b}
\bibinfo{author}{\bibfnamefont{A.~N.} \bibnamefont{Petrov}},
  \bibinfo{author}{\bibfnamefont{N.~S.} \bibnamefont{Mosyagin}},
  \bibnamefont{and} \bibinfo{author}{\bibfnamefont{A.~V.} \bibnamefont{Titov}},
  \bibinfo{journal}{Phys.\ Rev.\ A} \textbf{\bibinfo{volume}{79}},
  \bibinfo{pages}{012505} (\bibinfo{year}{2009}).

\bibitem[{\citenamefont{Skripnikov}(2017)}]{Skripnikov:17c}
\bibinfo{author}{\bibfnamefont{L.~V.} \bibnamefont{Skripnikov}},
  \bibinfo{journal}{J.\ Chem.\ Phys.} \textbf{\bibinfo{volume}{147}},
  \bibinfo{pages}{021101} (\bibinfo{year}{2017}).

\bibitem[{\citenamefont{Fleig}(2017)}]{Fleig:17}
\bibinfo{author}{\bibfnamefont{T.}~\bibnamefont{Fleig}},
  \bibinfo{journal}{Phys.\ Rev.\ A} \textbf{\bibinfo{volume}{96}},
  \bibinfo{pages}{040502(R)} (\bibinfo{year}{2017}).

\bibitem[{\citenamefont{Petrov}(2011)}]{Petrov:11}
\bibinfo{author}{\bibfnamefont{A.~N.} \bibnamefont{Petrov}},
  \bibinfo{journal}{Phys.\ Rev.\ A} \textbf{\bibinfo{volume}{83}},
  \bibinfo{pages}{024502} (\bibinfo{year}{2011}).

\bibitem[{\citenamefont{Cossel et~al.}(2012)\citenamefont{Cossel, Gresh,
  Sinclair, Coffey, Skripnikov, Petrov, Mosyagin, Titov, Field, Meyer
  et~al.}}]{Cossel:12}
\bibinfo{author}{\bibfnamefont{K.~C.} \bibnamefont{Cossel}},
  \bibinfo{author}{\bibfnamefont{D.~N.} \bibnamefont{Gresh}},
  \bibinfo{author}{\bibfnamefont{L.~C.} \bibnamefont{Sinclair}},
  \bibinfo{author}{\bibfnamefont{T.}~\bibnamefont{Coffey}},
  \bibinfo{author}{\bibfnamefont{L.~V.} \bibnamefont{Skripnikov}},
  \bibinfo{author}{\bibfnamefont{A.~N.} \bibnamefont{Petrov}},
  \bibinfo{author}{\bibfnamefont{N.~S.} \bibnamefont{Mosyagin}},
  \bibinfo{author}{\bibfnamefont{A.~V.} \bibnamefont{Titov}},
  \bibinfo{author}{\bibfnamefont{R.~W.} \bibnamefont{Field}},
  \bibinfo{author}{\bibfnamefont{E.~R.} \bibnamefont{Meyer}},
  \bibnamefont{et~al.}, \bibinfo{journal}{Chem.\ Phys.\ Lett.}
  \textbf{\bibinfo{volume}{546}}, \bibinfo{pages}{1 } (\bibinfo{year}{2012}).

\end{thebibliography}

\end{document}